\documentclass{desyproc}

\begin{document}
%------------------------------------
\title{Solar paraphotons}

%for single authors the superscripts are optional
\author{{\slshape Sergey Troitsky}\\[1ex]
Institute for Nuclear Research of the Russian Academy of Sciences,\\
60th October Anniversary Prospect 7A,
117312, Moscow, Russia}

% if the proceedings are available online (e.g. at Indico)
% please enter the contribution ID or file_name below for the DOI
%\contribID{32}
\contribID{troitsky\_sergey}

% TO THE CONFERENCE EDITORS:
% please update the following information
% before sending the template to the authors
% \confID{800}  % if the conference is on Indico uncomment this line
\desyproc{DESY-PROC-2011-04}
\acronym{Patras 2011} % if you want the Acronym in the page footer uncomment this line
\doi  % if there is an online version we will register DOIs

\maketitle

\begin{abstract}
I revisit the question of production of paraphotons, or hidden photons, in
the Sun and suggest that a simultaneous observations of solar flares by
conventional instruments and by axion helioscopes may provide a discovery
channel for paraphotons.
\end{abstract}

{\bf 1. Introduction.}
Hidden sectors, which interact
very weakly  with the observable world, are a usual
ingredient of theories extending the Standard Model and aimed at the
explanation of its parameters and their hierarchies.
Commonly, the interaction between the observable and
hidden sectors is mediated by a very heavy particle and appears in the
effective lagrangian, which describes the physics at the experimentally
testable energies, through non-renormalizable terms with couplings
suppressed by inverse powers of the mediator mass. It has been understood
long ago, however, that there generally exist renormalizable interactions
between two sectors, so-called portals, whose strength is not suppressed
by the mediator mass. Unless protected by some symmetries, these
interactions may be strong enough to allow tests of the hidden sector even
for the mediator masses of order the Planck scale. The Standard Model
fields allow for three kinds of such interactions: (i) the quartic
coupling of the Higgs scalar with some hidden scalar field (so-called
Higgs portal), (ii) the Yukawa coupling with neutrino, Higgs and a hidden
fermion, and (iii) the kinetic mixing term between the Standard-Model and
hidden $U(1)$ gauge fields. Here, we concentrate on the latter case,
first discussed in Ref.~\cite{Okun} where the gauge boson of the
additional $U(1)$ group was called a paraphoton. The kinetic mixing term,
which mixes the field strength of a hidden $U(1)$ gauge field with that of
the electromagnetic (or hypercharge) $U(1)$, is allowed by Lorenz and
gauge invariance and is renormalizable. Even if absent at the tree level,
it should be therefore generated by loop corrections unless a particular
symmetry prohibits it \cite{Kolda}.
There is no lack of theoretical models which have
sufficient freedom to justify observable paraphotons with almost arbitrary
parameters allowed by experimental constraints. Some part of the
paraphoton parameter space (which for our purposes consists of the
paraphoton mass $m$ and the kinetic-mixing coupling $\chi$ but in general
includes also the gauge coupling of the hidden $U(1)$),
however, have
special phenomenological importance because these values have been invoked
for models explaining either experimental anomalies or the origin of the
Standard-Model parameters. We emphasize three particularly interesting
regions.

(1) Mimicking extra neutrinos in the CMB. Recent cosmological results
suggest  that the effective number $N_{\rm eff}$ of light
neutrino species is larger than three: the WMAP7 data \cite{WMAP-Neff}
gives $N_{\rm eff} =4.34^{+0.86}_{-0.88}$, in agreement with somewhat less
precise SDSS Data Release 7 \cite{SDSS-Neff} and Atacama Cosmology
Project \cite{Atacama-Neff}. It has been suggested that paraphotons with
mass $m$ in the range $(10^{-5} \div 10^{-2})$~eV may mimic extra neutrino
species, the change in $N_{\rm eff}$ determined \cite{hiddenCMB} by the
mixing
$\chi$. For the WMAP7 values quoted above, $\chi= (1.1\div 2.4) \times
10^{-6}$.

(2) String compactifications with TeV-scale gravity. Some of popular
approaches to the gauge hierarchy problem in the Standard Model imply
lowering the fundamental gravitational scale down to the values of order
electroweak scale or slightly higher. This is usually achieved in models
with extra space dimensions, in particular, in string models. Paraphotons
are generic by-product in these compactification models. In a certain
class of the latter, the fundamental string scale is related
\cite{Ringwald-string} to the kinetic-mixing parameter . The string scale
is bounded from below by the early LHC results to be larger than a few
TeV; its values within $(5\div 1000)$~TeV would correspond to
$\chi\sim (10^{-12} \div 10^{-10})$ for a wide range of possible paraphoton
masses.

(3) ``Unified'' or ``secluded'' dark matter and hidden SM Higgs. These
approaches attempt to explain the anomalies observed by DAMA, PAMELA and
INTEGRAL, as well as possible non-observation of a light ($\sim 100$~GeV)
Higgs boson with unusual decay channels. Though quite different, all these
scenarios point to $\chi\sim (10^{-4} \div 10^{-3})$ and paraphoton mass in
the GeV range.

For different values of the parameters $m$ and $\chi$, various
experimental techniques have been implemented to search for a potential
signal of paraphotons. None was found, resulting in severe limits on the
parameter space, see e.g.\ Ref.~\cite{para-limits-review} for a review.

In the Sun, paraphotons may be produced from solar thermal
photons by means of the kinetic mixing, see e.g.\ Refs.~\cite{Popov1,
Redondo, RedondoZur, STToAppear}. The oscillation probability for the most
general case will be presented elsewhere ~\cite{STToAppear}; here we will
be interested in two limiting cases important for the Sun, namely the case
of optically thick emission region (the solar interior) and that of
transparent emission region (solar outer atmosphere and solar flares).

\begin{figure}[t]
\centering
\begin{tabular}{m{0.48\columnwidth}m{0.48\columnwidth}}
\noindent
\begin{tabular}{c}
\includegraphics[width=0.4\columnwidth]{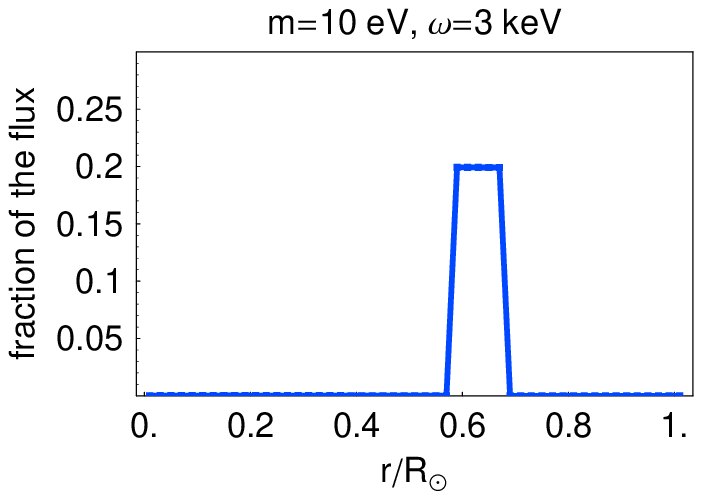}\\
\includegraphics[width=0.4\columnwidth]{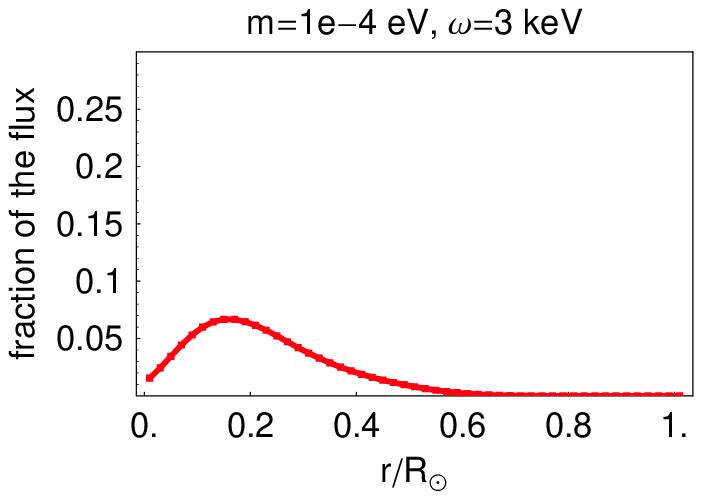}
\end{tabular}&
\noindent
\includegraphics[width=0.47\columnwidth,origin=lb]{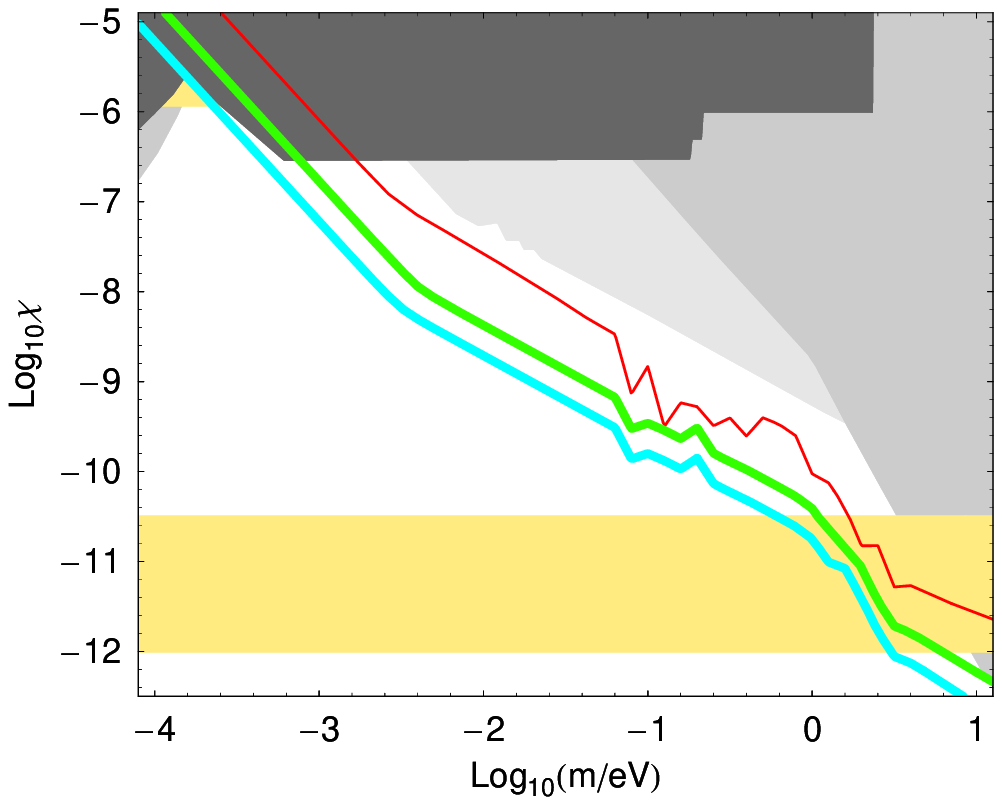}
\end{tabular}
%\centering
%\begin{minipage}{0.5\columnwidth}
%\includegraphics[width=\columnwidth]{slicem1E3000.eps}\\
%\includegraphics[width=\columnwidth]{sliceM4E3000.eps}
%\end{minipage}~%
%\includegraphics[width=0.5\columnwidth]{excl.eps}
\caption{
\label{fig:slice}
\textit{Left:} Normalized contributions of various parts of the Sun to the
total paraphoton flux. Upper panel: no resonance, the central part
dominates. Lower panel: resonance, a thin slice dominates.
\textit{Right:} Paraphoton parameters. Dark gray:
laboratory exclusion; light gray: astrophysical exclusion; very light
gray: CAST exclusion from Ref.~\cite{Redondo}. Very light gray (yellow
online): theoretically favoured regions (see the Introduction). It is
expected that future X-ray helioscopes will exclude the space above lines
(top to bottom, the planned CAST upgrade and two options for the
next-generation axion helioscope IAXO \cite{NGAH}).}
\end{figure}

{\bf 2. Contribution of the optically thick Sun.} The Sun has a rather
sharp boundary where the density, and hence the transparency, changes by
many orders of magnitude. It has been shown (see, e.g.,
Ref.~\cite{Redondo}) that for paraphotons of keV energies, the
contribution
of the optically thick interior dominates. The total flux of paraphotons
in this case is given by \cite{Redondo}
\[
\frac{d\Phi}{d\omega}=
\frac{3\times 10^{24}}{{\rm cm}^{2}\cdot {\rm s}\cdot {\rm eV}}
\left(\frac{\chi}{10^{-5}} \right)^{2}
\left(\frac{m}{{\rm eV}}   \right)^{4}
f_{1}(\omega,m),
\]
where
\[
f_{1}(\omega,m) =  1~{\rm eV} \times
\omega^{2}\int\limits_{0}^{1}\! d\xi \, \xi^{2} \frac{\Gamma(\xi
R_{\odot})}{{\rm e}^{\omega/T(\xi R_{\odot})}-1}\,
\frac{1}{\left(\omega_{p}^{2}(\xi R_{\odot}) -m_{1}^{2}  \right)^{2}
+\omega^{2} \Gamma(\xi R_{\odot})^{2}}.
\]
$m$ and $\omega$ are the paraphoton mass and energy, $\omega_{p}$ is
the usual plasma frequency, $\chi$ is the mixing coupling $\xi$ is the
radial coordinate measured in the units of the solar radius, $R_{\odot}$,
while $T$ and $\Gamma$ determine the temperature and the inverse mean free
path of a photon with energy $\omega$, calculated at a given point in the
Sun, respectively. The plasma frequency in the Sun varies roughly from
0.1~eV to 300~eV and for $m$ within this range, the integral is saturated
by a contribution of a rather thin resonance slice, the paraphotosphere;
otherwise, high-temperature inner parts dominate (see
Fig.~\ref{fig:slice}, left panel).
The right panel of Fig.~\ref{fig:slice}
gives approximate exclusion limits on the plane
of paraphoton parameters (mass and coupling) expected for future X-ray
helioscopes, together with current experimental bounds.

{\bf 3. Solar flares.}
We assume that a flare happens in a small region with constant temperature
and electron density and its emission is thermal. Then the ratio of the
paraphoton flux from the flare to the photon flux at the same energy is
approximately
$P/(1-P)$, where $P$ is
the probability of the conversion at the emission point. The duration of
the flare is $\sim 10^{3}$~s and normally, since $P\ll 1$,  only a tiny
number of paraphotons reach the detector during this time. The
situation changes drastically if the density of plasma in the flare
happens to be such that the resonance takes place. Then $P\approx 1/2$ and
a large number of photons were converted to paraphotons. It is easy to
demonstrate that for a detector with area $S$ and background noize $n$,
the 95\% CL exclusion limit on the mixing parameter $\chi$ for one
particular resonant mass may be determined by
$$
\chi \gtrsim  8 \times 10^{-7}
\left(\frac{F_{\mbox{obs}}}{10^{5}~{\rm
cm}^{\mbox{-2}}\, {\sf s}^{-1} \,{\rm
eV}^{\mbox{--1}}} \right)^{-1/2}
\left(\frac{t}{1~{\rm s}}
\right)^{-1/4}
\left(\frac{n}{10^{-3}~{\rm Hz}}
\right)^{1/4}
\left(\frac{S}{10~{\rm cm}^{2}}
\right)^{-1/2}
\left(\frac{\omega}{\rm keV}
\right)^{-1/2},
$$
where $F_{\mbox{\rm{obs}}}$ is the flux of solar \textit{photons} with
energy $\omega$ detected from the flare. In this formula, $t$ is the
period of time when the flare keeps resonant conditions, that is the
plasma frequency does not change more than the resonance width. We note
that while a flare looks as a rapid process at a particular wavelength,
the change in appearence is related mostly to change in the temperature
and not in the density (see e.g.\ discussion in Ref.~\cite{Aschw} and
particular numbers in Ref.~\cite{flare-numbers}), so that $t \gtrsim 1$~s
is a typical value. However, in practice this approach can hardly be used
to constrain $\chi$ because one does not know the conditions in the flare
with the required precision.

Instead, I suggest that the observation of flares with helioscopes may be
a discovery channel for paraphotons. Indeed, the resonant conversion of
photons to paraphotons in the flare manifests itself not only by
appearence of paraphotons but also by a $\sim 1/2$ drop in the regular
\textit{photon} flux. The study of light curves with excellent time
resolution is possible with various instruments (e.g., SOXS, SPHINX etc.).
One may search them for temporal coincidences between events in
helioscopes and drops in the flare lightcurves; the background is almost
zero and even a single coincident event may serve as a strong evidence for
paraphotons.

{\bf 4. Acknowledgements.}
I acknowledge collaboration with Dmitry Gorbunov, Martyn Davenport and
Konstantin Zioutas at various stages of this work. I am indebted to the
organizers of the Mykonos workshop for the invitation to this very
interesting meeting and to many participants of the meeting, notably to
Javier Redondo, for illuminating discussions. This work was supported in
part by
the
 grants RFBR 10-02-01406, 11-02-01528 and FASI 02.740.11.0244, by the
 grant of the President of the Russian Federation NS-5525.2010.2 and by
 the Dynasty foundation.

\end{document}